\documentclass[11pt]{article}

\usepackage[utf8]{inputenc}
\usepackage[english]{babel}
\usepackage[margin=2.3cm]{geometry} %
\usepackage{graphicx} %
\usepackage{amsthm, amsmath, amssymb} 
\usepackage{tcolorbox} 
\usepackage{placeins} %
\usepackage[numbers]{natbib} %
\usepackage{lineno}	
\usepackage{authblk} %
\usepackage{appendix} %
\usepackage{mathrsfs}  %
\usepackage{authblk} %

\usepackage{booktabs} %

\usepackage{setspace} %

\usepackage[version=4]{mhchem}

\usepackage{pdflscape}   %
\usepackage{url} %
\newcommand{\papertitle}{Contact point geometry governs structural build-up at rest in Portland cement-limestone blends}

\newcommand{\storagemodulus}{G'}
\newcommand{\storagemodulusref}{\storagemodulus_0}
\newcommand{\storagemodulusnorm}{\tilde{G}}

\newcommand{\cumulheat}{H}
\newcommand{\cumulheatJ}{Q}

\newcommand{\cumulheatzero}{H_{0}}

\newcommand{\deltacumulheatHzero}{\cumulheat-\cumulheatzero}

\newcommand{\deltacumulheatJ}{\Delta \cumulheatJ}

\newcommand{\volume}{V}
\newcommand{\volumeproducts}{\volume_{\mathrm{prod}}}
\newcommand{\changevolumeproducts}{\Delta \volumeproducts}

\newcommand{\nuprod}{\nu}

\newcommand{\productsthickness}{h}

\newcommand{\deltaproductsthickness}{\Delta \productsthickness}

\newcommand{\sem}{\mathrm{SEM}}
\newcommand{\psd}{\mathrm{PSD}}
\newcommand{\ssa}{\mathrm{SSA}}
\newcommand{\surface}{S}
\newcommand{\surfaceofgrain}{\surface_{\mathrm{grain}}}
\newcommand{\volumeofgrain}{\volume_{\mathrm{grain}}}

\newcommand{\mass}{m}

\newcommand{\characteristiclength}{l_c}
\newcommand{\density}{\rho}

\newcommand{\interparticleinteraction}{\sigma}
\newcommand{\numcontacts}{N}

\newcommand{\solidvolfrac}{\phi}

\newcommand{\ie}{{\textit{i.e.}}}
\newcommand{\eg}{{\textit{e.g.}}}

\newcommand{\tricalciumsilicate}{\ce{C3S}}
\newcommand{\tricalciumsilicatestoichiometric}{\ce{Ca3SiO5}}
\newcommand{\portlandite}{\ce{CH}}
\newcommand{\portlanditestoichiometric}{\ce{Ca(OH)2}}
\newcommand{\watercementnotation}{\ce{H}}
\newcommand{\water}{\ce{H2O}}
\newcommand{\csh}{\ce{C-S-H}}
\newcommand{\cshpointsixtyseven}{\ce{C_{1.67}SH_{2.1}}}
\newcommand{\cshstoichiometric}{\ce{Ca_{1.67}SiO2(OH)_{3.33}.{0.43}H2O}}

\newcommand{\molarmass}{\mathrm{M}}

\newcommand{\enthalpy}{H^0}
\newcommand{\deltaenthalpy}{\Delta \enthalpy}
\newcommand{\enthaplyformation}{\enthalpy_f}
\newcommand{\deltaenthaplyformation}{\Delta \enthaplyformation}

\newcommand{\deltaenthaplyreaction}{\Delta \enthalpy_{\mathrm{reaction}}}
\newcommand{\deltaenthaplyformationproducts}{\Delta \enthalpy_{f,\mathrm{products}}}
\newcommand{\deltaenthaplyformationreactants}{\Delta \enthalpy_{f,\mathrm{reactants}}}

\newcommand{\soichiometriccoeffproducts}{n_p}
\newcommand{\soichiometriccoeffreactants}{n_r} \graphicspath{ {./} }
\begin{document}

\title{\papertitle}
\author[1]{Luca Michel}
\author[1]{Antoine Sanner}
\author[2]{Franco Zunino}
\author[1]{Robert J. Flatt}
\author[1]{David S. Kammer}
\affil[1]{Institute for Building Materials, ETH Zurich, Switzerland}
\affil[2]{Department of Civil and Environmental Engineering, University of California at Berkeley, USA.} 
\maketitle

\section*{Abstract}

The early stiffening of fresh cement paste plays a key role in shaping and stability during casting and 3D printing. In Portland cement systems, this phenomenon arises from the formation of calcium–silicate–hydrate ($\csh$), which stiffens grain-to-grain contacts. However, the role of powder characteristics such as particle size and morphology remains poorly understood. Here, we vary fineness and grain shape by blending Portland cement with either coarse or fine limestone, leveraging the affinity of $\csh$ to nucleate on limestone surfaces. By coupling calorimetry and rheometry, we relate the amount of formed hydration products to the increase in stiffness, and show that the mechanism of contact stiffening through $\csh$ formation remains unchanged with limestone addition. Nevertheless, the rate of stiffening varies across blends. We find that these rates correlate with a characteristic length scale that captures particle size and shape. These results demonstrate that early stiffening depends not only on the amount of hydration products formed, but also on the geometry of the contacts where these products form, offering a framework for understanding more complex systems such as limestone–calcined clay cements.

\section{Introduction}
On the microscopic level, fresh cement paste forms a dense, flocculated network of grains suspended in water. This network arises from attractive interactions between cement particles~\cite{flattDispersionForcesCement2004, bonacciContactMacroscopicAgeing2020} and is progressively reinforced by the formation of hydration products~\cite{rousselOriginsThixotropyFresh2012, michelStructuralBuildrestInduction2024}. The combined effect of these interparticle forces and hydration-induced strengthening leads to a gradual increase in macroscopic mechanical properties over time, a phenomenon commonly known as structural build-up or structuration~\cite{rousselOriginsThixotropyFresh2012, michelStructuralBuildrestInduction2024, reiterRoleEarlyAge2018a}.

Experimentally, structural build-up is characterized by evolutions of storage modulus or yield stress~\cite{yuanMeasurementEvolutionStructural2017, reiterStructuralBuildupDigital2019a, flalo2025117929, jiaoThixotropicStructuralBuildup2021, bellottoCementPastePrior2013a}. In the past years, \emph{time} evolutions have been widely used for structural build-up investigations. While such approaches are insightful for practical applications, such as 3D printing with concrete~\cite{perrot3DPrintingConcrete2019, perrotPredictionLateralForm2015}, they do not directly link hydration reactions to changes in macroscopic mechanical properties. Moreover, time-resolved data are influenced by hydration kinetics, limiting their ability to provide fundamental insights~\cite{hanEffectMixingMethod2015}. A recent approach combining rheometry and calorimetry has addressed this limitation by directly correlating hydration reactions with the evolution of storage modulus~\cite{michelStructuralBuildrestInduction2024}. However, it has so far been applied only to Portland cement paste with a fixed particle size distribution and specific surface area. The impact of variations in these powder characteristics on build-up at rest remains unknown, yet it can be expected to play a significant role in practical systems, such as blended cements.

Blending Portland cement with other mineral powders is a well-established strategy to reduce the clinker content in cementitious materials, thus lowering their carbon footprint~\cite{aitcinSupplementaryCementitiousMaterials2016, lothenbachSupplementaryCementitiousMaterials2011}. Among various formulations, Limestone Calcined Clay Cement (LC3) stands out, allowing up to 50\% clinker replacement~\cite{francozuninoLimestoneCalcinedClay2021}. LC3 includes Portland cement, limestone, and calcined clay, creating a complex system in which drastic changes in structural build-up and workability have been observed~\cite{michelEarlyageWorkabilityLoss2023, moghulFlowLossSuperplasticized2024}. These effects are often attributed to calcined clay, but the combined addition of limestone and clay makes it difficult to isolate the influence of powder fineness and grain morphology.

In this context, Portland cement-limestone blends offer a simpler, canonical model for understanding how variations in particle size distribution, specific surface area, and grain geometry affect structural build-up at rest. Since these factors are also present in LC3, studying cement-limestone systems provides a controlled framework that captures key physical aspects of more complex blends, while excluding potential chemical effects introduced by reactive clays. This lays a foundation for systematically investigating the role of powder morphology in the early mechanical evolution of multi-component systems, like LC3.

In this work, we investigate how adding limestone to Portland cement impacts structural build-up at rest, focusing on variations in powder fineness and grain geometry. Coupled calorimetry and rheometry measurements allow us to express the evolution of storage modulus as a function of the average thickness of hydration products forming on grain surfaces. Based on these results, we find that the general build-up mechanism remains unchanged compared to plain Portland cement, namely that contact points are strengthened by $\csh$ formation. We observe that finer powders lead to a faster increase in storage modulus for a given hydration product thickness. These differences are canceled out when accounting for grain morphology, whereby data collapse onto a single master curve when plotting the normalized storage modulus against a normalized thickness of products. Hence, this suggests that the geometry of grain-to-grain contacts plays a key role in structural build-up at rest. Our findings indicate that structural build-up at rest is governed not only by ``how much'' hydration products form, but also ``into what geometry'' these products form. This highlights the critical role that contact point geometry plays in structural build-up at rest.

\section{Materials}

\subsection{Raw powders}
In this work, we investigate the effect of limestone addition and powder fineness on structural build-up in fresh Portland cement paste at rest. The cement used is a Portland cement CEM I 52.5R (Holcim Normo 5R), blended with either a coarse limestone (Sigma-Aldrich Calcium Carbonate, 99.95\% dry basis) or a fine limestone (Omya Betocarb UF, high-purity calcium carbonate). The specific surface area ($\ssa$) of the Portland cement, coarse limestone, and fine limestone is 1.1, 0.56, and 5.53 m$^2$/g, respectively. 
$\ssa$ measurements were performed using a BET multi-point nitrogen physisorption apparatus (Micromeritics Tristar II 3020).

\begin{figure}[htbp]
    \centering
    \includegraphics[width=0.55\linewidth]{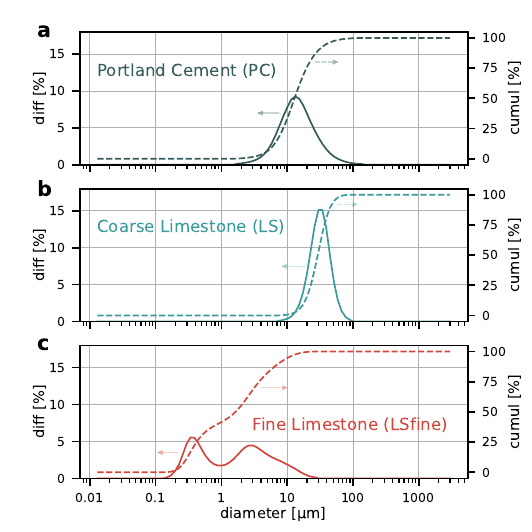}
    \caption{\textit{Particle size distributions of the raw powders measured by laser diffraction. \textbf{a)} Portland cement. \textbf{b)} Coarse limestone. \textbf{c)} Fine limestone. The solid lines represent the differential size distributions (left axes), and the dashed lines display the cumulative size distributions (right axes).}}
    \label{fig:p02:psd_raw_powders}
\end{figure}

Particle size distributions ($\psd$) were determined via laser diffraction using a HORIBA Partica LA-950 Laser Scattering Particle Size Distribution Analyzer. Before measuring  the final size distribution, the particles were dispersed by ultrasound three times to ensure proper dispersion. The particles are dispersed in isopropanol without use of dispersant. The coarse limestone has a particle size distribution comparable to that of Portland cement, whereas the fine limestone exhibits a distribution shifted toward smaller diameters, as shown in Fig.~\ref{fig:p02:psd_raw_powders}.

\subsection{Blends investigated}

\begin{figure}[htbp]
    \centering
    \includegraphics[width=0.6\linewidth]{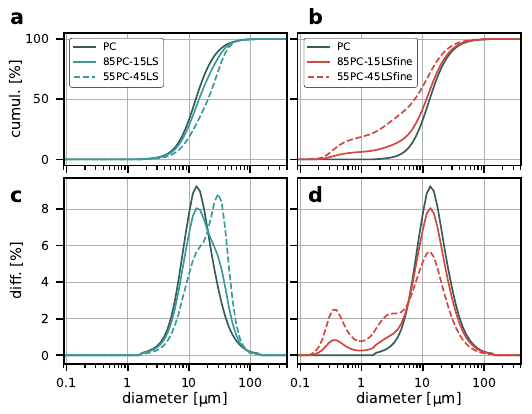}
    \caption{\textit{Particle size distributions of the blends investigated. \textbf{a,b)} Cumulative size distributions for Portland cement-coarse limestone blends, and Portland cement-fine limestone blends, respectively. \textbf{c,d)} Differential size distributions for Portland cement-coarse limestone blends, and Portland cement-fine limestone blends, respectively. The size distributions of Portland cement are systematically reported as reference.}}
    \label{fig:p02:psd_blends}
\end{figure}

We add Limestone (both coarse and fine) to Portland cement in fractions of 15\% and 45\% by mass. The selected water-to-binder (w/b) ratios ensure that the pastes are homogeneously mixed while avoiding stability issues such as sedimentation and bleeding. An overview of the blends investigated is given in Table~\ref{tab:p02:compositions}. The particle size distributions of the blends are reported in Fig.~\ref{fig:p02:psd_blends}, where, compared to Portland cement, the addition of coarse limestone shifts the size distributions towards larger diameters, and the addition of fine limestone results in distributions containing larger fractions of fine particles.

\begin{table}[htbp]
\caption{\textit{Compositions and w/b ratios of the investigated blends. The system ID is used to label the different mixes throughout the manuscript. PC = Portland cement. LS = Coarse limestone. LSfine = Fine limestone. We report the lower bound w/b ratio used during mixing to minimize self-mixing energy effects~\cite{mantellatoRelatingEarlyHydration2019}.}}
\label{tab:p02:compositions}
\centering
\begin{tabular}{|c|c|c|c|}
\hline
\textbf{System ID} & \textbf{Composition in mass \%} & \textbf{w/b ratios}  & \textbf{lower bound w/b} \\ \hline
100 PC	& 100\% Portland cement  & 0.36, 0.38,  &0.34\\
		&					& 0.4, 0.42	&	\\ \hline
85PC - 15LS & 85\% Portland cement  & 0.38, 0.4,  	&0.32\\
 		     &    15\% Coarse Limestone  & 0.44	 & \\ \hline
55PC - 45LS   & 55\% Portland cement   & 0.48, 0.55, 0.59 & 0.48\\ 
			& 45\% Coarse Limestone &	& \\ \hline
85PC - 15LSfine  & 85\% Portland cement  & 0.4 (2x), 0.45	&0.4\\ 
			   & 15\% Fine Limestone	  & 			& \\ \hline
55PC - 45LSfine  & 55\% Portland cement  & 0.4 (2x), 0.45	&0.4 \\ 
			& 45\% Fine Limestone	 & 			& \\ \hline

\end{tabular}
\end{table}

\subsection{Characterizing powder morphology}
\label{sec:p02:characterizing_powder_morphology}

\begin{figure}[htbp]
    \centering
    \includegraphics[width=0.95\linewidth]{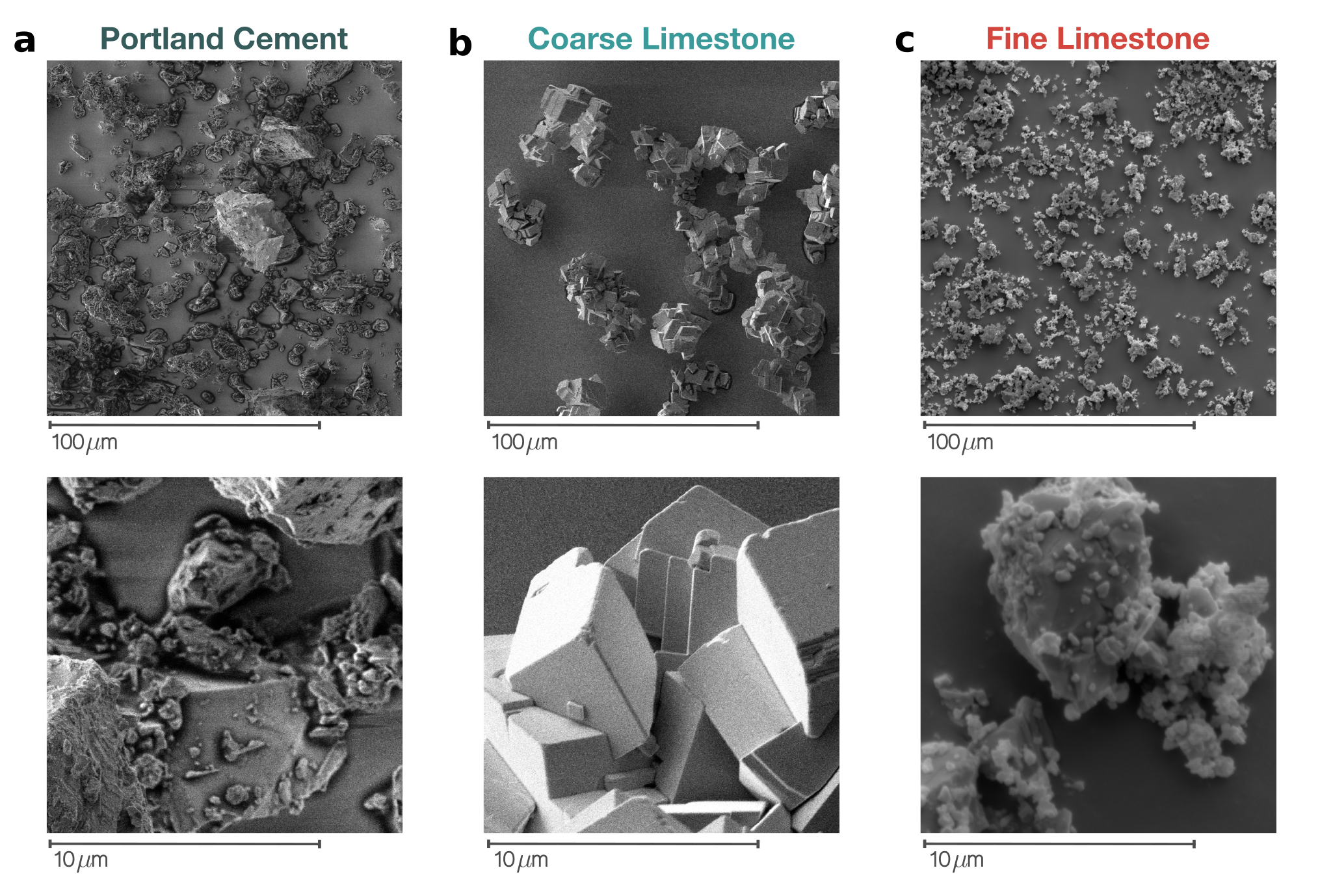}
    \caption{\textit{Scanning electron microscope images of the raw powders. \textbf{a)} Portland cement. \textbf{b)} Coarse limestone. \textbf{c)} Fine limestone.} }
    \label{fig:p02:sem_raw_powders}
\end{figure}

A direct way to observe particle morphology at the micrometer scale is by means of scanning electron microscopy ($\sem$) images, as reported in Fig~\ref{fig:p02:sem_raw_powders}. From these images, we observe that the Portland cement exhibits a heterogeneous and angular morphology, characteristic of ground powders, see Fig.~\ref{fig:p02:sem_raw_powders}a. In contrast, the coarse limestone features a highly cubic morphology typical of calcite, see Fig.~\ref{fig:p02:sem_raw_powders}b. This morphology results from the manufacturing process, whereby no grinding was done after the precipitation of the limestone. The fine limestone, however, has been ground during the manufacturing process, and thus appears rounder and visually more similar to Portland cement, see Fig.~\ref{fig:p02:sem_raw_powders}c. The bimodal nature of the fine limestone seen in the $\psd$ measurements in Fig.~\ref{fig:p02:psd_raw_powders}c also appears in Fig.~\ref{fig:p02:sem_raw_powders}c, where fine particles are seen to cover coarser grains.

The differences in grain shape observed in the $\sem$ images in Fig.~\ref{fig:p02:sem_raw_powders} imply that blending Portland cement with varying amounts of coarse and fine limestone will result in variations in overall powder morphology. These variations are expected to have an impact on structural build-up at rest, resulting in a need to quantify variations in powder morphology, beyond qualitative statements based on $\sem$ images.

Given information about the volume and surface area of a powder, the most straightforward way to characterize powder morphology, albeit simplistic, is to compute the associated characteristic length $\characteristiclength$, defined by the volume-to-surface ratio as
\begin{equation}
    \characteristiclength = \frac{\volume_{\mathrm{powder}}}{\surface_{\mathrm{powder}}} = \frac{1}{\density\,\ssa} ~,
\label{eq:p02:lc_rho_ssa}
\end{equation}
where $\density$ is the powder density and $\ssa$ its specific surface area. This formulation provides a quantitative means of assessing the differences in morphology between the powders investigated, which will be used in our analysis of structural build-up at rest. We report values $\characteristiclength$ for the raw powders, as well as for each blend investigated, in Tab.~\ref{tab:p02:lc_values}. The addition of coarse limestone results in increased $\characteristiclength$, and the addition of fine limestone reduces the resulting $\characteristiclength$ values.

\begin{table}[htbp]
\caption{\textit{Characteristic length scale $\characteristiclength$ corresponding to the volume-to-surface ratio for the raw powders and blends investigated.}}
\label{tab:p02:lc_values}
\centering
\begin{tabular}{|c|c|c|c|}
\hline
\textbf{Composition} & \textbf{density} $\density$ [g/cm$^3$] & \textbf{SSA} [m$^2$/g] & $\characteristiclength = 1/(\density\,\ssa)$ [nm]  \\ \hline
100 PC	& 3.15 & 1.09 & 291 \\ \hline
100 LS	& 2.7 & 0.56 & 661  \\ \hline
100 LSfine	& 2.7 & 5.53 & 67  \\ \hline
85PC - 15LS & 3.08 & 1.01 & 321 \\ \hline
55PC - 45LS   & 2.95 & 0.85 & 398  \\ \hline
85PC - 15LSfine  & 3.08 & 1.76 & 185  \\ \hline
55PC - 45LSfine  & 2.95 & 3.09 & 110  \\ \hline

\end{tabular}
\end{table}

\section{Structural build-up investigation}
\subsection{Coupled calorimetry and rheometry}

We investigate structural build-up at rest by coupling isothermal calorimetry and small amplitude oscillatory shear rheology measurements to track how the storage modulus evolves with the formation of hydration products. The experimental procedure used to simultaneously record both quantities on two instances of the same sample follows the approach established in our previous work on plain Portland cement~\cite{michelStructuralBuildrestInduction2024}.

We mix each sample with a 4-bladed propeller stirrer mounted on an IKA Eurostar EURO-ST P CV mixer. We first add the dry powder to a fraction of the total mixing water corresponding to a lower bound w/b ratio (reported in Table~\ref{tab:p02:compositions}) and mix the sample for one minute at $500~\mathrm{RPM}$. We then add the remaining mixing water needed to reach the targeted w/b ratios and mix the paste for two more minutes at 500 RPM. This staggered procedure minimizes the influence of self-mixing energy effects, as proposed by \citet{mantellatoRelatingEarlyHydration2019}. The heat released by hydration reactions is monitored with a TAM AIR isothermal calorimeter at $23\,^\circ\mathrm{C}$ with glass vials filled with around $5$~g of paste. 

We perform rotational rheometry measurements on $10~\mathrm{g}$ of paste with an Anton Paar Physica MCR 501 rheometer in parallel plate geometry using serrated plates with a $1~\mathrm{mm}$ gap. The diameters of the upper and lower plates are $25~\mathrm{mm}$ and $50~\mathrm{mm}$, respectively. We start the rheometer measurement directly after the start of the calorimeter measurement. We use an Anton Paar PTD 200 Peltier device to set the temperature in the rheometer at $23\,^\circ\mathrm{C}$, and a plastic hood limits drying of the sample over the measurement period. The measurement is strain-controlled, except for a stress-controlled step at the liquid-solid transition of the sample~\cite{ovarlezInfluenceShearStress2008a}, directly after a pre-shear event. The rheometer protocol is composed of a large amplitude oscillatory shear (LAOS) pre-shear of $30~\mathrm{seconds}$ at an amplitude of $10~\mathrm{\%}$ and frequency of $1~\mathrm{Hz}$, followed by a stress-controlled rest period where a zero shear stress $\tau=0$ is applied for $30~\mathrm{seconds}$. 

\begin{figure}[htbp]
    \centering
    \includegraphics[width=0.7\linewidth]{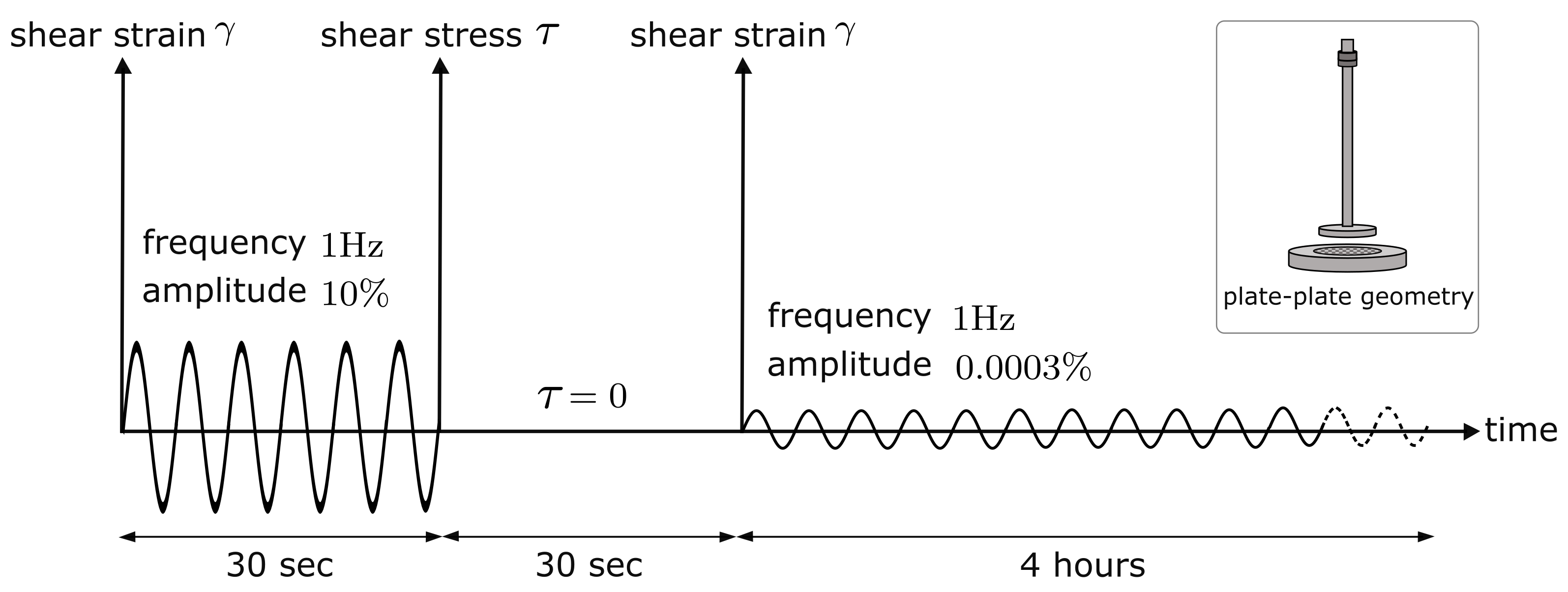}
    \caption{\textit{Overview of the rheometer protocol.} }
    \label{fig:p02:rheoprot}
\end{figure}

\begin{figure}[htbp]
    \centering
    \includegraphics[width=0.65\linewidth]{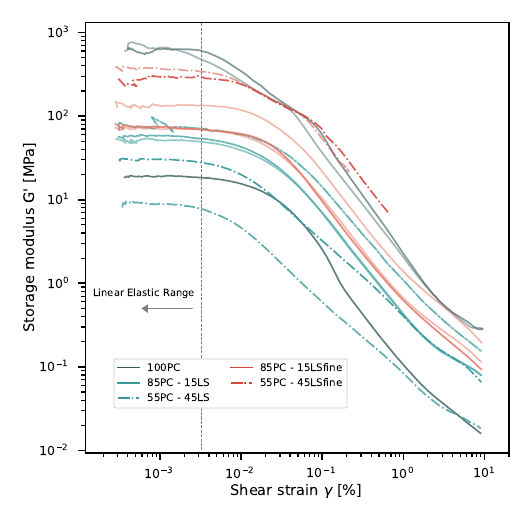}
    \caption{\textit{Amplitude sweeps to assess the linear elastic range of the samples. Higher transparency curves correspond to lower w/b ratios. The detailed w/b ratio values are given in Table~\ref{tab:p02:compositions}. } }
    \label{fig:p02:ler}
\end{figure}

After these steps, we consider the sample to be in a homogeneous state. A small amplitude oscillatory shear (SAOS) measurement is then started to record the storage modulus evolution over time. The SAOS measurement is performed at an amplitude of $0.0003~\mathrm{\%}$ and frequency of $1~\mathrm{Hz}$. We give a graphical overview of the rheometer protocol in Fig.~\ref{fig:p02:rheoprot}. In Portland cement pastes, as well as in blends of Portland cement and \emph{coarse} limestone, the storage modulus measurement lasts for $4~\mathrm{hours}$. In blends of Portland cement and \emph{fine} limestone, the storage modulus measurement only lasts $2.5~\mathrm{hours}$ due to the faster kinetics resulting from the increased powder fineness, which limits the measurement window over time. The amplitude of the SAOS measurement is set such that the samples are probed inside their linear elastic range, which we assess through amplitude sweeps, as reported in Fig.~\ref{fig:p02:ler}. All the storage modulus data are post-processed by applying a low-pass filter with cutoff~0.012.

\subsection{Time evolutions of heat and storage modulus}

Isothermal calorimetry data allow us to track the advancement of hydration reactions in time. From heat rate data, we identify characteristic hydration regimes such as the initial peak, induction period, and acceleration period. Due to noise introduced during sample insertion into the calorimeter, we discard the first hour of data~\cite{michelStructuralBuildrestInduction2024}. As a result, our measurements primarily probe the induction and the early acceleration periods of hydration. We measure systems at different w/b ratios, which results in different hydration kinetics across the samples due to self-mixing energy effects~\cite{mantellatoRelatingEarlyHydration2019, juillandEffectMixingEarly2012}. This can be seen in Fig.~\ref{fig:p02:timeevols}a, where for a given system, heat rate curves are shifted in time. To account for these variations in hydration kinetics, we use the onset of the acceleration period as a reference state, following our previous work~\cite{michelStructuralBuildrestInduction2024}. We identify the onset by the characteristic ``bump'' in heat rate, likely associated with Portlandite nucleation~\cite{damidotHydrationDilutedStirred1994}, as highlighted in Fig.~\ref{fig:p02:timeevols}a. We note here that this definition of the onset, and thus of reference state, holds in the \emph{absence} of admixtures. In presence of admixtures, the respective position of the bump will depend on their type and dosage, making the definition of a reference state more complex. The cumulative heat quantifies the extent of hydration that has occurred over a given time interval. We track changes in cumulative heat relative to the onset of the acceleration period, $\deltacumulheatHzero$, as reported in Fig.~\ref{fig:p02:timeevols}b, providing a reference-based measure of the hydration progress~\cite{michelStructuralBuildrestInduction2024}. While it has not been proven that the bump is an actual signature of the onset of the acceleration period, it remains a feature consistently appearing through our samples allowing us to align cumulative heat evolutions on a well-defined reference point.

The storage modulus $\storagemodulus$ quantifies the stiffness of a paste under shear deformations~\cite{schultzUseOscillatoryShear1993a}. In cementitious systems, it increases over time due to the hydration reactions strengthening the network of suspended grains. The evolution of $\storagemodulus$ is affected by the w/b ratio and blend composition, as shown in Fig.~\ref{fig:p02:timeevols}c. Lower w/b ratios (denser systems) lead to systematically faster $\storagemodulus$ evolutions, which can be attributed to the higher number of contact points per unit volume and faster hydration kinetics~\cite{michelStructuralBuildrestInduction2024}. Blend composition also impacts the time evolution of $\storagemodulus$. The addition of coarse limestone tendentially slows down $\storagemodulus$ development compared to plain Portland cement. In contrast, adding \emph{fine} limestone systematically accelerates the evolution of $\storagemodulus$, leading to faster kinetics, as seen, \eg, in the $\storagemodulus$ values at 2.5~$\mathrm{h}$ for w/b = 0.4 in Fig.~\ref{fig:p02:timeevols}c. We keep track of the storage modulus at the onset of the acceleration period, $\storagemodulusref$, which serves as a reference storage modulus.

\begin{landscape}
\thispagestyle{empty} %
\begin{figure}[htbp]
    \centering
    \includegraphics[width=0.98\linewidth]{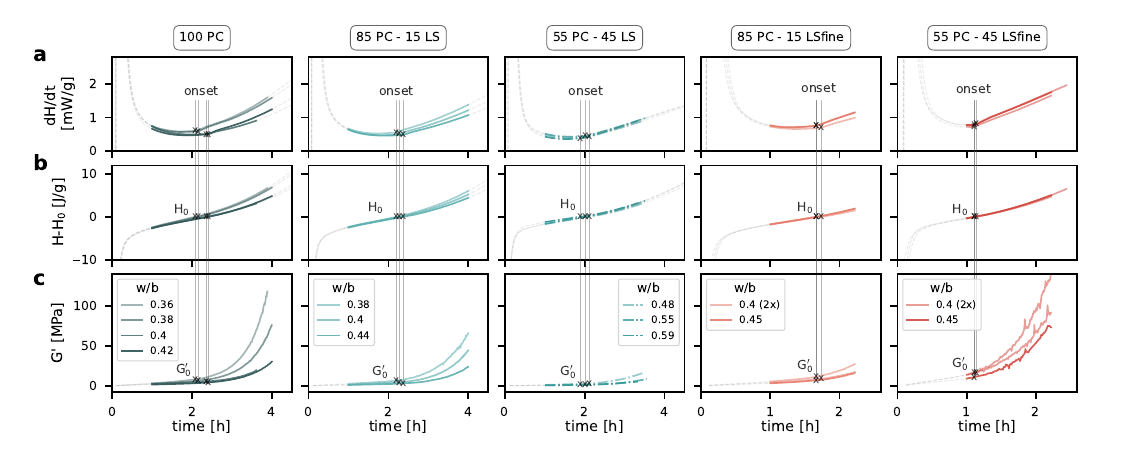}
    \caption{\textit{ Time evolutions of hydration heat and storage modulus in Portland cement-limestone blends. \textbf{a) Heat rate} evolution over time for the different blends investigated. The first hour of heat rate data is discarded due to noise from sample insertion in the calorimeter. The onset of the acceleration period is identified at the bump in heat rate. \textbf{b) Changes in cumulative heat} computed as the time integral of heat rate. $\cumulheatzero$ is the cumulative heat at the onset of the acceleration period. \textbf{c) Storage modulus} evolutions at different w/b ratios measured by small amplitude oscillatory shear. $\storagemodulusref$ is the storage modulus at the onset of the acceleration period. The first hour of $\storagemodulus$ data is discarded as it cannot be coupled to noisy heat data. For PC-LSfine systems, w/b ratio 0.4 is measured twice (labeled ``2x'' in the respective legends).}}
    \label{fig:p02:timeevols}
\end{figure}
    
\end{landscape}

\section{Hydration products forming on the surface of grains}
\label{p02:sec:hydration_products_surface_grains}

The next step in our investigation consists in directly linking $\storagemodulus$ evolutions to the advancement of hydration. Previous studies have shown that the evolution of $\storagemodulus$ is driven by the amount of hydration products forming on the surface of cement grains~\cite{rousselOriginsThixotropyFresh2012, michelStructuralBuildrestInduction2024, reiterStructuralBuildupDigital2019a}. However, considering the \emph{total} volume of products formed does not explicitly account for powder fineness. Given that powder fineness is a parameter varied in this study, it must be properly accounted for. Indeed, finer systems, characterized by higher $\ssa$, provide a greater surface area for hydration products to form. As a result, for a given volume of products, the amount deposited per grain is lower in finer systems. To properly account for this effect, it is essential to consider how hydration products distribute across the available surface area of the grains.

The distribution of products on the surfaces of grains depends on the type of hydration reaction taking place. In Portland cement pastes, within the experimental window considered here, the main hydration reaction is alite hydration~\cite{marchonMechanismsCementHydration2016}, whereby $\csh$ forms on the surface of the grains. We assume that the same process takes place in Portland cement-limestone blends, given that limestone has been reported to act as a template for $\csh$ formation~\cite{berodierImpactSupplementaryCementitious, berodierUnderstandingFillerEffect2014}. Thus, in the systems studied here, we consider uniform $\csh$ deposition across \textit{all} available surfaces. 

\begin{figure}[htbp]
    \centering
    \includegraphics[width=0.5\linewidth]{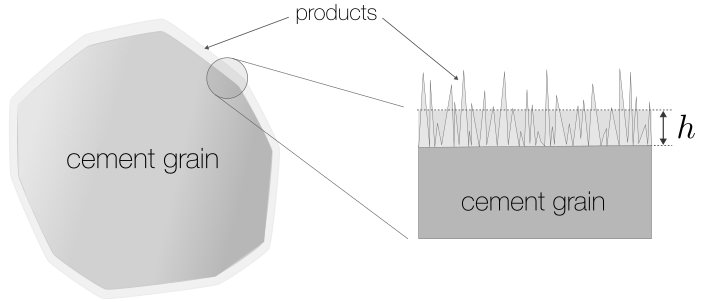}
    \caption{\textit{Modeling $\csh$ growth as a layer of thickness $\productsthickness$.} }
    \label{fig:p02:products_layer}
\end{figure}

Over the timescales investigated, $\csh$ forms in a needle-like morphology~\cite{berodierImpactSupplementaryCementitious, scrivenerAdvancesUnderstandingCement2019, ouziaNeedleModel_New2019}. However, as a first order approximation due to lack of detailed spatial information about the needle growth, we model $\csh$ formation as a growing layer of thickness $\productsthickness$, as illustrated in Fig.~\ref{fig:p02:products_layer}.
 
With the assumption of uniform distribution of $\csh$ on all grain surfaces, and assuming a layer of products infinitesimally thin compared to the grain size~\cite{zuninoInfluenceSulfateAddition2022}, we estimate the thickness $\productsthickness$ of the hydration product layer based on the total volume of products formed, $\volumeproducts$, and the total surface area of the grains, $\surfaceofgrain$, through
\begin{equation}
    \productsthickness = \frac{\volumeproducts}{\surfaceofgrain} ~.
    \label{eq:p02:thickness_Vprod_Sgrain}
\end{equation}

To compute the volume of hydration products formed, we consider alite hydration and the formation of Jennite-type $\csh$~\cite{matscheiThermodynamicPropertiesPortland2007}, as given by:
\begin{equation}
\tricalciumsilicate + 3.43\watercementnotation \ce{->} \cshpointsixtyseven + 1.33\portlandite ~.
    \label{eq:p02:c3s_hydration}
\end{equation}
We note that the computation of $\productsthickness$ does not rely on the exact composition of $\csh$, and thus, we proceed with Jennite type $\csh$ given that thermodynamic data are available for this phase~\cite{matscheiThermodynamicPropertiesPortland2007}. 

Given the molar masses, densities, enthalpies of formation, and initial composition of the paste, we compute the change in volume of hydration products, $\changevolumeproducts$, corresponding to a given change in cumulative heat, $\deltacumulheatJ$, as
\begin{equation}
    \changevolumeproducts = \nuprod \deltacumulheatJ ~,
    \label{eq:p02:DV_nu_DH}
\end{equation}
where $\nuprod = 4.1\cdot 10^{-4}$ $\mathrm{cm^3/J}$ assuming the reaction in Eq.~\ref{eq:p02:c3s_hydration}, and $\deltacumulheatJ$ is expressed in Joules. The calculation of the coefficient $\nuprod$ is provided in Appendix~\ref{p02:appendix:enthalpies_and_volume_of_products}.

Within the time frame considered here, $\csh$ formation does not significantly alter the morphology of the cement grains. Instead, the grains dissolve via etch pit formation~\cite{nicoleauDiTricalciumSilicate2013, juillandDissolutionTheoryApplied2010, juillandAdvancesDissolutionUnderstanding2017}, which does not substantially change their specific surface area. Consequently, we assume that $\surfaceofgrain$ remains constant throughout our experimental window. This allows us to compute the change in product thickness, $\deltaproductsthickness$, by combining Eqs.~\ref{eq:p02:thickness_Vprod_Sgrain} and~\ref{eq:p02:DV_nu_DH} into
\begin{equation}
    \deltaproductsthickness = \frac{\changevolumeproducts}{\surfaceofgrain} 
    = \frac{\nuprod(\deltacumulheatHzero)}{\ssa} ~,
    \label{eq:p02:thicknesschange_cumulheat_ssa}
\end{equation}
where $\deltacumulheatHzero$ is the change in cumulative heat recorded by isothermal calorimetry, expressed in Joules per unit mass of binder, and $\ssa$ is the specific surface area of the powder, expressed as a surface area per unit mass of binder. 

$\deltaproductsthickness$ provides a more physically meaningful perspective on hydration advancement compared to quantities like changes in cumulative heat directly obtained from isothermal calorimetry measurements. In the following section, we will examine how $\storagemodulus$ evolves as the hydration-products layer grows, aiming for a mechanistic understanding of the link between the stiffness evolution and the growth of hydration products.

\section{Storage modulus vs. thickness of hydration products}

\begin{figure}[htbp]
    \centering
    \includegraphics[width=0.6\linewidth]{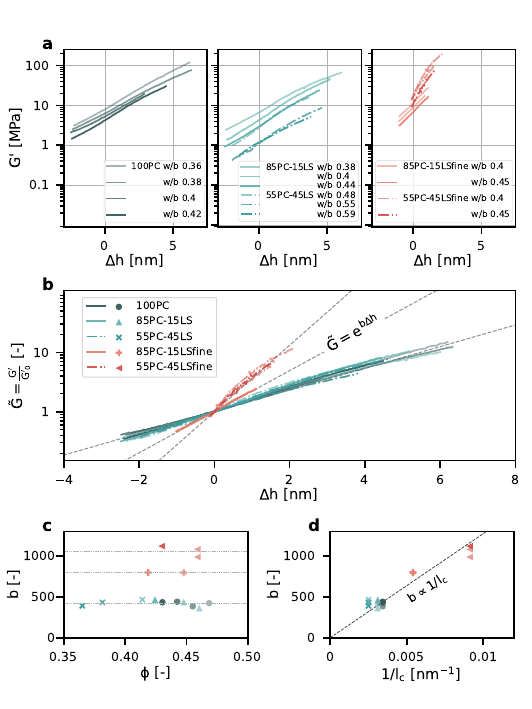}
    \caption{\textit{Limestone addition affects the stiffness evolution compared to plain Portland cement paste. \textbf{a)} Storage modulus evolution with changes in thickness of hydration products at different w/b ratios. Left: Portland cement. Middle: blends of Portland cement with coarse limestone. Right: blends of Portland cement with fine limestone. \textbf{b)} Normalized storage modulus vs. changes in thickness of hydration products for blends with and without limestone. \textbf{c)} Exponent of the $\storagemodulusnorm$ vs. $\deltaproductsthickness$ relation as a function of the initial solid volume fraction $\solidvolfrac$ (unhydrated grains). \textbf{d)} The exponent of the $\storagemodulusnorm$ vs. $\deltaproductsthickness$ relation scales linearly with the inverse of the characteristic length, $1/\characteristiclength$.} }
    \label{fig:p02:G_dh}
\end{figure}

We now examine how $\storagemodulus$ evolves with $\deltaproductsthickness$ across the different blends investigated. For each system investigated, $\storagemodulus$ systematically exhibits an exponential increase with $\deltaproductsthickness$, as shown in Fig.~\ref{fig:p02:G_dh}a. This dependence can be expressed as
\begin{equation} \label{eq:p02:G_aexp_bdh}
\storagemodulus = ae^{b\deltaproductsthickness} ~,
\end{equation}
whereby exponential fits on the $\storagemodulus$ vs. $\deltaproductsthickness$ data consistently yield $R^2$ values above 98\%, as detailed in Appendix~\ref{sec:p02:appendix_expfits_R2}. For a given mix composition, the prefactor $a$ in Eq.~\ref{eq:p02:G_aexp_bdh} varies with w/b ratio, as can be seen from the varying intercepts at $\deltaproductsthickness=0$ in Fig.~\ref{fig:p02:G_dh}a. In contrast, the exponent $b$ in Eq.~\ref{eq:p02:G_aexp_bdh} is independent of w/b ratio \emph{for a given mix composition}, resulting in the constant slopes observed in Fig.~\ref{fig:p02:G_dh}a. Overall, we observe that denser systems result in stiffer evolutions, and that for a fixed mix composition, the exponent, and thus the nature of the exponential evolution, remains unchanged.

Normalizing the $\storagemodulus$ vs. $\deltaproductsthickness$ evolutions by the storage modulus at the onset of the acceleration period $\storagemodulusref$ (defined in Fig.~\ref{fig:p02:timeevols}c) cancels out the dependence of $\storagemodulus$ on w/b ratio, as reported in Fig.~\ref{fig:p02:G_dh}b. We report the dependence of $\storagemodulusref$ on $\solidvolfrac$ in Appendix~\ref{sec:p02:appendix_G0phi}. For a given mix composition, all curves collapse on a single evolution, irrespective of w/b ratio. Defining the normalized storage modulus $\storagemodulusnorm=\storagemodulus/\storagemodulusref$, the evolutions in Fig.~\ref{fig:p02:G_dh}b can be expressed as 
\begin{equation} \label{eq:p02:Gtilde_exp_bdh}
\storagemodulusnorm = e^{b\deltaproductsthickness} ~,
\end{equation}
whereby the exponent $b$ remains independent of w/b ratio after the normalization by $\storagemodulusref$. This observation is further illustrated in Fig.~\ref{fig:p02:G_dh}c, where $b$ is plotted against the initial solid volume fraction $\phi$, \ie, the volume fraction of unhydrated cement grains. Interestingly, $b$ has the same value for Portland cement paste as well as for blends of Portland cement with coarse limestone, irrespective of the coarse limestone content. In contrast, addition of fine limestone to Portland cement results in an increase of $b$.

We further investigate the dependence of $b$ on mix composition by taking into account the powder morphology, characterized by the characteristic length $\characteristiclength$ defined in Sec.~\ref{sec:p02:characterizing_powder_morphology}. To this end, we plot $b$ against $1/\characteristiclength$, as reported in Fig.~\ref{fig:p02:G_dh}d. This reveals a linear relation between $b$ and $1/\characteristiclength$, suggesting that powder morphology has a direct impact on the exponential increase of $\storagemodulus$ with $\deltaproductsthickness$.

\begin{figure}[htbp]
    \centering
    \includegraphics[width=0.8\linewidth]{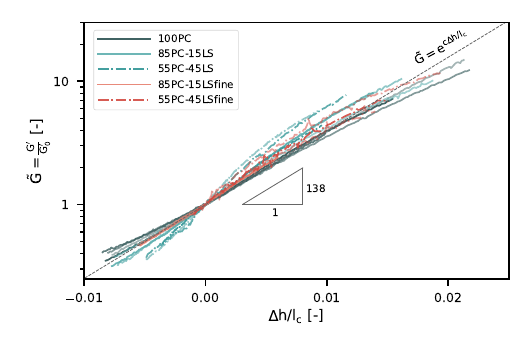}
    \caption{\textit{Taking the grain morphology into account for structural build-up investigations. Normalized storage modulus $\storagemodulusnorm$ with respect to change in products thickness normalized by the characteristic length $\deltaproductsthickness/\characteristiclength$. Lighter colors correspond to lower w/b ratios, w/b ratio values are provided in Fig.~\ref{fig:p02:G_dh}}a.}
    \label{fig:p02:G_dhoverlc}
\end{figure}

Based on the observed linear relation between $b$ and $1/\characteristiclength$, we plot $\storagemodulusnorm$ evolutions against a normalized thickness of hydration products $\deltaproductsthickness/\characteristiclength$, as reported in Fig.~\ref{fig:p02:G_dhoverlc}. This collapses all the curves on a single master curve, independent of w/b ratio and mix composition. We note that some systems with 45\% coarse limestone deviate a bit from the unique exponential. We nevertheless consider this deviation of second order compared to the overall good overlap of all different systems, and will discuss it further in Sec.~\ref{sec:p02:discussion}. The resulting exponential relation can be described by 
\begin{equation} \label{eq:p02:Gtilde_exp_cdhlc}
\storagemodulusnorm = e^{c\deltaproductsthickness/\characteristiclength} ~,
\end{equation}
where the exponent $c$ has a value $138\pm 21$.

\section{Discussion}
\label{sec:p02:discussion}

In Portland cement pastes, structural build-up at rest is governed by the progressive strengthening of grain-to-grain contacts due to the formation of $\csh$ during the induction and acceleration periods of hydration~\cite{michelStructuralBuildrestInduction2024}. The storage modulus evolution in this regime can be expressed as
\begin{equation}
\storagemodulus \propto f(\numcontacts) \cdot \interparticleinteraction(\cumulheat) ~,
\label{eq:p02:GfNsigmaH}
\end{equation}
where $f(\numcontacts)$ is a function of the total number of grain-to-grain contacts $\numcontacts$, which is independent of the contact stiffness $\interparticleinteraction$~\cite{bonacciContactMacroscopicAgeing2020, zhouYieldStressConcentrated1999, zhouChemicalPhysicalControl2001, flattYodelYieldStress2006}. Given the low amounts of products formed at very early age~\cite{flattDispersionForcesCement2004}, \ie, over the time periods relevant to structural build-up, $\numcontacts$ remains unchanged, such that the hydration $\cumulheat$ only affects the grain-to-grain stiffness $\interparticleinteraction$. Normalizing the storage modulus evolution by a reference storage modulus $\storagemodulusref$ cancels out the effect of $\numcontacts$ in Eq.~\ref{eq:p02:GfNsigmaH}, isolating the impact of hydration on contact stiffness
\begin{equation}
    \storagemodulusnorm = \frac{\storagemodulus}{\storagemodulusref} = \frac{\interparticleinteraction(\cumulheat)}{\interparticleinteraction(\cumulheatzero)} ~,
    \label{eq:p02:Gnorm_sigmaH_sigmaH0}
\end{equation}
where $\cumulheatzero$ represents a reference hydration state, which can be taken, for instance at the onset of the acceleration period. Thus, evolutions of the normalized storage modulus $\storagemodulusnorm$ allow to cancel out the effect of w/b ratio, as it only plays a role in $f(\numcontacts)$, without impacting the grain-to-grain stiffness $\interparticleinteraction$. This has been reported for Portland cement~\cite{michelStructuralBuildrestInduction2024}, and our data suggest that this also holds for Portland cement-limestone blends. Indeed, evolutions of the normalized storage modulus are independent of w/b ratio for a given powder fineness, see Fig.~\ref{fig:p02:G_dh}b. This implies that the addition of limestone does not alter the basic mechanism underlying structural build-up at rest, namely a strengthening of contact points by hydration products.

In this work, we study the effect of powder fineness on build-up at rest by considering both a coarse and a fine limestone. To account for powder fineness, we consider the thickness of hydration products forming on the surface of the grains, $\deltaproductsthickness$, rather than the total volume of products forming. However, our results reveal that this is not sufficient to fully capture the trends observed experimentally. Specifically, addition of fine limestone results in variations in normalized storage modulus evolutions that cannot be explained solely by changes in $\deltaproductsthickness$, suggesting that other microstructural features play a role.

Incorporating both powder morphology and fineness allows us to capture all the trends observed experimentally. To this end, we consider the volume-to-surface ratio of the powders as a characteristic length $\characteristiclength$ capturing differences in powder morphology. Expressing $\storagemodulusnorm$ against $\deltaproductsthickness/\characteristiclength$, which is a normalized products thickness,  collapses all curves on a single master curve, see Fig.~\ref{fig:p02:G_dhoverlc}. This suggests that $\characteristiclength$ effectively captures the relevant geometrical features of the contact points strengthened by $\csh$, and is thus key to a mechanistic understanding of structural build-up at rest.

\begin{figure}[htbp]
    \centering
    \includegraphics[width=0.8\linewidth]{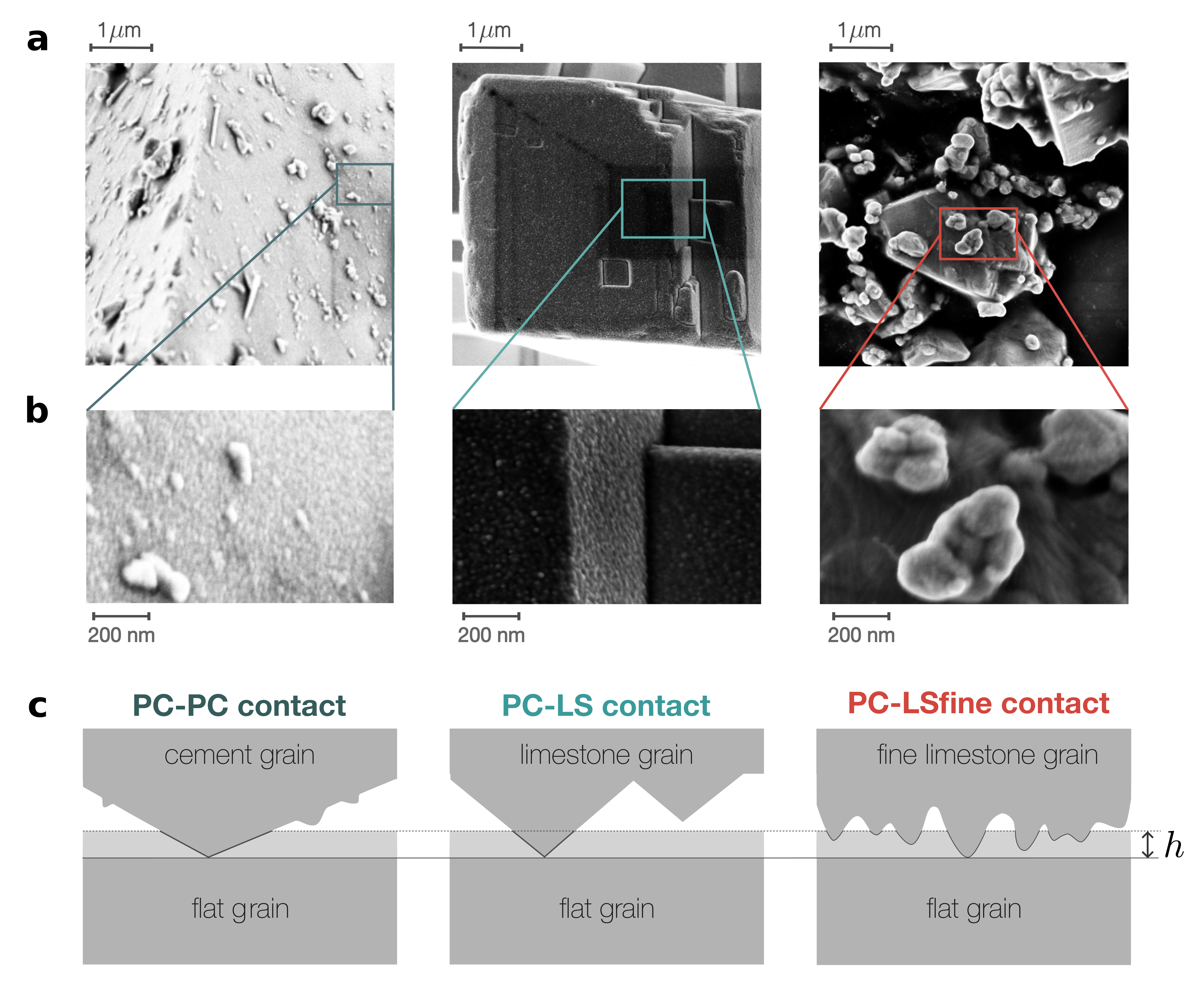}
    \caption{\textit{Scanning electron microscope images of the powders at comparatively high magnification. From left to right: Portland cement, coarse limestone, and fine limestone. \textbf{a)} Grain morphology at the micrometer scale. \textbf{b)} Zoom-in around features at a scale comparable to $\characteristiclength$. \textbf{c)} Schematic illustration highlighting the differences in contact geometry based on the different grain morphologies. For a given thickness of hydration products, rougher contacts result in more bridged surface. The hydrates are represented by the light gray layer, shown only on the bottom surface to avoid overloading the schematic. For considerations involving the surface bridged by hydration products, this representation is mathematically equivalent to having the hydration products symmetrically distributed on both surfaces.} }
    \label{fig:p02:contact_illustration}
\end{figure}

For the blends investigated here, $\characteristiclength$ ranges from $100$ to $400~\mathrm{nm}$, as reported in Tab.~\ref{tab:p02:lc_values}. SEM images taken at a scale comparable to $\characteristiclength$ reveal distinct morphological differences between the fine limestone and the other powders, see Fig.~\ref{fig:p02:contact_illustration}a\&b. At that scale, the Portland cement and coarse limestone feature rather flat surfaces and angular morphologies. In contrast, the fine limestone displays a very round and rough morphology. The increased roughness of the fine limestone arises from its bimodal nature, where fine grains adsorbing on the surface of larger grains result in asperities of size comparable to $\characteristiclength$. Based on these observations, we draw schematic grain-to-grain contacts for Portland cement paste, as well as blends containing coarse and fine limestone, see Fig.~\ref{fig:p02:contact_illustration}c. For a given thickness of hydration products, it appears that a rougher contact results in more area bridged by $\csh$, which can qualitatively explain why in Fig~\ref{fig:p02:G_dh}b, for a given $\deltaproductsthickness$, systems containing fine limestone present a steeper increase in $\storagemodulusnorm$.

While $\storagemodulusnorm$ vs. $\deltaproductsthickness/\characteristiclength$ evolutions collapse all data on a single master curve, some systems containing high fractions of coarse limestone appear to deviate from the general trend in Fig.~\ref{fig:p02:G_dhoverlc}. Looking at the overall powder morphology in Fig.~\ref{fig:p02:sem_raw_powders}, the coarse limestone stands out with its extremely cubic grain geometry. While the use of this cubic limestone allows to vary powder morphology in the design of experiment, the quantification of powder morphology is done based on a single scalar value, $\characteristiclength$. A single scalar value is likely not enough to fully capture morphological features of size distributed powders, like those considered here. Nevertheless, the volume-to-surface ratio is one of the few quantities embedding powder morphology that can be obtained from direct independent measurements, in this case density and specific surface area. More advanced morphological characterization would likely require direct analysis from $\sem$ images, where the broad size distribution of the powders considered makes it a challenging task. Hence, while $\characteristiclength$ likely does not capture all the relevant features, it appears as an accessible and relatively robust means of assessing powder morphology, embedding aspects relevant to structural build-up at rest. 

Based on the qualitative observations in Fig.~\ref{fig:p02:contact_illustration}, it appears that not only the amount of hydration products matters for structural build-up, but also the geometry of the contacts in which these products grow. Regarding the scale of the relevant geometrical features of the contacts, it is likely dependent on the size of products forming. Indeed, the size range of $\characteristiclength$ is comparable to the length of $\csh$ needles growing on the surface of the grains~\cite{zuninoInfluenceSulfateAddition2022}. From a mechanistic standpoint, this suggests that the effective area bridged by $\csh$ plays a key role in dictating structural build-up. As illustrated in Fig.\ref{fig:p02:contact_illustration}c, for a given $\deltaproductsthickness$, a rougher contact geometry (as found in contacts involving fine limestone, with a large surface curvature) results in a larger bridged surface, leading to a stronger increase in storage modulus. \citet{rousselOriginsThixotropyFresh2012} suggested that the size or number of $\csh$ bridges drives the increase in macroscopic elastic modulus. Our data provide experimental evidence supporting this mechanism. However, our results highlight that solely considering $\csh$ growth is not sufficient to fully capture build-up at rest. Additionally, our findings are consistent with contact mechanics theories, where an increase in effective contact area leads to enhanced stiffness~\cite{barberContactMechanics2018}.

A crucial implication of our findings is that plotting the storage modulus evolution against cumulative heat release per gram of binder ($\cumulheat$) is fundamentally equivalent to plotting it against the product thickness normalized by the characteristic length ($\deltaproductsthickness/\characteristiclength$). This is because cumulative heat release per gram of binder scales with the volume of hydration products formed per volume of binder, which can be expressed as
\begin{equation}
\deltacumulheatHzero \propto \frac{\changevolumeproducts}{\volumeofgrain} = \frac{\deltaproductsthickness\cdot\surfaceofgrain}{\volumeofgrain} = \frac{\deltaproductsthickness}{\characteristiclength} ~,
\end{equation}
where $\characteristiclength = \volumeofgrain/\surfaceofgrain$. This demonstrates that the $\storagemodulusnorm$ vs. $\deltacumulheatHzero$ curves inherently capture contact geometry effects, reinforcing the robustness of this approach for describing build-up behavior across different systems. However, $\storagemodulusnorm$ vs. $\deltacumulheatHzero$ hide details required for fundamental understanding, which only appear upon explicit consideration of $\storagemodulusnorm$ vs. $\deltaproductsthickness/\characteristiclength$ evolutions.

We modeled $\csh$ growth as a layer of thickness $\productsthickness$ fully covering the surface of the grains. To this end, we used the dry $\ssa$ of the powders. However, shortly after contact with water, the $\ssa$ of Portland cement increases substantially due to the formation of early hydration products. For the Portland cement used here, after 10 minutes of hydration, the $\ssa$ increases from 1.1~m$^2$/g to 2.2~m$^2$/g~\cite{mantellatoRelatingEarlyHydration2019}. Which $\ssa$ value is used in the analysis impacts the surface available for $\csh$ growth, whereby the extra surface after contact with water probably comes from aluminate hydrates, which may not play a role for $\csh$ deposition. Using the $\ssa$ after contact with water over the dry $\ssa$ changes the $\storagemodulus$~vs.~$\deltaproductsthickness$ evolutions, as detailed in Appendix~\ref{sec:p02:appendix_wet_ssa}. However, the $\storagemodulusnorm$~vs.~$\deltaproductsthickness/\characteristiclength$ evolutions are independent of the chosen $\ssa$ value, as it is canceled out in the $\deltaproductsthickness/\characteristiclength$ normalization (see Eqs.~\ref{eq:p02:thicknesschange_cumulheat_ssa} and~\ref{eq:p02:lc_rho_ssa}):
\begin{equation}
\deltaproductsthickness\cdot \frac{1}{\characteristiclength} = \frac{\nuprod(\deltacumulheatHzero)}{\ssa} \cdot \density\ssa = \nuprod(\deltacumulheatHzero)\density ~.
\end{equation}
Thus, the general statement that both the amount of products as well as the geometry of contacts in which these products grow holds independently of the exact contact geometry, which is apriori unknown.

The influence of other hydration products forming at very early age such as ettringite~\cite{marchonMechanismsCementHydration2016}, cannot be investigated with the current experimental setup. Indeed, due to noise in the calorimetry data coming from sample insertion, coupling of $\storagemodulus$ and $\deltaproductsthickness$ can only start after the first hour of reaction~\cite{michelStructuralBuildrestInduction2024}. Nevertheless, the contributions of products forming during the first hour of reaction are most probably canceled out in the $\storagemodulus/\storagemodulusref$ normalization. Furthermore, our measurements focus on the late induction and early acceleration periods of hydration, where the $\storagemodulusnorm$~vs.~$\deltaproductsthickness/\characteristiclength$ evolutions are mostly driven by $\csh$ formation. We expect this to hold as long as the system does not become undersulfated, with an inversion of the silicate peak with the sulfate depletion. Our heat flow curves do not suggest that our systems become undersulfated. It is however not clear what the consequences would be with respect to our analysis in cases where such undersulfation may occur.

While this study supports the idea that contact geometry plays a key role in structural build-up, it is important to note that our study focuses on Portland cement-limestone blends, where limestone primarily acts as a template for $\csh$ formation~\cite{berodierImpactSupplementaryCementitious, berodierUnderstandingFillerEffect2014}. This means that, in these systems, hydration product nucleation and growth remain comparable to those observed in Portland cement alone. However, when considering other supplementary cementitious materials (SCMs) such as calcined clays, additional factors may influence build-up. Specifically, SCMs with different chemical compositions and surface chemistries could alter the affinity of $\csh$ to nucleate and grow on their surfaces, potentially modifying the relationship between hydration product formation, contact reinforcement, and stiffness evolution. As a result, while our results provide strong evidence for the role of contact geometry in structural build-up, further investigations are needed to assess whether the same scaling laws hold for systems with chemically distinct SCMs.

Beyond its fundamental implications, our findings may also have practical applications in optimizing the rheological properties of cementitious suspensions. Understanding how contact morphology influences build-up could inform mix design strategies aimed at tailoring early stiffness evolution, with potential benefits for construction processes such as 3D printing~\cite{reiterRoleEarlyAge2018a}, formwork pressure~\cite{manuallyaddedOvarlezPhysical2007}, and workability retention~\cite{moghulFlowLossSuperplasticized2024}.

\section{Conclusion}

In this study, we showed that limestone addition does not alter the fundamental mechanism of structural build-up at rest compared to plain Portland cement paste, namely a strengthening of grain-to-grain contacts through $\csh$ formation. While blending powders with different finenesses results in varying thicknesses of hydration products, this alone is not sufficient to fully describe build-up at rest. It is crucial to consider powder morphology, which we successfully capture through a characteristic length computed as the volume-to-surface ratio of the powders. This suggests that contact geometry plays an important role in structural build-up, emphasizing that it is not only ``how much'' hydration products form, but also ``into what geometry'' they form that matters. While our results are robust for Portland cement-limestone blends, further work is needed to assess whether the same scaling holds for systems containing chemically distinct supplementary cementitious materials, such as calcined clays. Our findings support the idea that contact area is a key determinant of structural build-up, with implications for both fundamental understanding and practical applications in cementitious systems.

\section*{CRediT authorship contribution statement}
\textbf{Luca Michel}: Methodology, Investigation, Formal Analysis, Data Curation, Visualisation, Writing -- Original Draft.\\
\textbf{Antoine Sanner}: Formal Analysis, Writing -- Review and Editing.\\
\textbf{Franco Zunino}: Resources, Writing -- Review and Editing.\\
\textbf{Robert J. Flatt}: Conceptualisation, Resources, Writing -- Review and Editing.\\
\textbf{David S. Kammer}: Conceptualisation, Supervision, Formal Analysis, Writing -- Review and Editing, Project administration, Funding acquisition.

\section*{Declaration of competing interest}
The authors declare that they have no known competing financial interests or personal relationships that could have appeared to influence the work reported in this paper.

\section*{Acknowledgements}
We thank Christian Franck, Nicolas Bain, Mohit Pundir, and Flavio Lorez for useful discussions during the manuscript preparation. We also thank Arturo Chao Correas for feedback on the manuscript draft. We gratefully acknowledge the Swiss National Science Foundation for financial support under grant number 200021\_200343. We also thank the Scientific Center for Optical and Electron Microscopy (ScopeM) of ETH Zurich for providing assistance with the generation of the scanning electron microscope images.

\section*{Data availability}
The data that support the findings of this study are openly available in ETH Research Collection at \url{https://doi.org/10.3929/ethz-b-000741976}, reference number ethz-b-000741976. The codes to generate the figures are available on gitlab: \url{https://gitlab.ethz.ch/cmbm-public/papers-supp-info/2025/buildup_pcls}.

\appendix
\section{Enthalpies of reaction and volume of products}
\label{p02:appendix:enthalpies_and_volume_of_products}

This appendix gives the detailed calculation of the parameter $\nuprod$ introduced in Eq.~\ref{eq:p02:DV_nu_DH}, allowing to compute the volume of products formed from cumulative heat data recorded by isothermal calorimetry. We consider alite hydration, described by
\begin{equation}
\ce{C3S + 3.43H -> C_{1.67}SH_{2.1} + 1.33CH} ~.
\label{eq:p02:appendix_alite_hydration}
\end{equation}

The physical and chemical properties of the compounds in equation~\ref{eq:p02:appendix_alite_hydration} are summarized in Table~\ref{tab:p02:physical_chemical_properties_appendix}.

\begin{table}[htbp]
\caption{Properties of the compounds considered in Eq.~\ref{eq:p02:appendix_alite_hydration}. Standard enthalpy of formation ($\Delta H^0_f$) values are taken from~\cite{matscheiThermodynamicPropertiesPortland2007}, and density values are taken from~\cite{matscheiThermodynamicPropertiesPortland2007, scrivenerPracticalGuideMicrostructural2018}.}
\label{tab:p02:physical_chemical_properties_appendix}
\centering
\begin{tabular}{@{}lcccc@{}}
\toprule
 & $\tricalciumsilicate$ & $\watercementnotation$ & $\cshpointsixtyseven$ & $\portlandite$ \\
 & \footnotesize{$\tricalciumsilicatestoichiometric$} & \scriptsize{$\water$} & \scriptsize{$\cshstoichiometric$} & \scriptsize{$\portlanditestoichiometric$} \\[0.9ex]
\midrule
$\molarmass$ $\left[\frac{g}{mol}\right]$ & 228 & 18 & 191.32 & 74 \\[1.3ex]
$\density$ $\left[\frac{g}{cm^3}\right]$ & 3.12 & 1 & 2.45 & 2.24 \\[1.3ex]
$\deltaenthaplyformation$ $\left[\frac{kJ}{mol}\right]$ & -2931 & -286 & -2723 & -985 \\[1.3ex]
$\mass_{\mathrm{initial}} \ \left[\frac{g}{g_{\mathrm{binder}}}\right]$ & 1 & w/c & 0 & 0\\
\bottomrule
\end{tabular}
\end{table}

The enthalpy of reaction $\deltaenthaplyreaction$ can be calculated with Hess's law~\cite{hessRecherchesThermochimiquesBulletin1840}:
\begin{equation}
\deltaenthaplyreaction = \Sigma \soichiometriccoeffproducts \deltaenthaplyformationproducts - \Sigma \soichiometriccoeffreactants \deltaenthaplyformationreactants ~,
\label{eq:hess_law}
\end{equation}
where $\deltaenthaplyformationproducts$, $\deltaenthaplyformationreactants$, $\soichiometriccoeffproducts$, and $\soichiometriccoeffreactants$ are the enthalpies of formation and stoichiometric coefficients of the products and reactants, respectively.

The enthalpy of formation of 1 mole of $\csh$ is then
\begin{equation*}
\deltaenthaplyreaction = \left[ 1\cdot \deltaenthalpy_{f,\cshpointsixtyseven} + 1.33\cdot \deltaenthalpy_{f,\portlandite} \right]- \left[ 1\cdot \deltaenthalpy_{f,\tricalciumsilicate} + 3.43\cdot \deltaenthalpy_{f,\watercementnotation} \right] = -121.1  \ \left[\frac{kJ}{mol}\right] ~.
\end{equation*}

The volume of products formed for one mole of reaction is obtained from the molar masses and densities of the reactants and products, along with their stoichiometric coefficients
\begin{equation}
\volumeproducts = \frac{\molarmass_{\cshpointsixtyseven}}{\density_{\cshpointsixtyseven}} + 1.33\cdot \frac{\molarmass_{\portlandite}}{\density_{\portlandite}} - \frac{\molarmass_{\tricalciumsilicate}}{\density_{\tricalciumsilicate}} = 48.9 \ \left[\frac{cm^3}{mol} \right] ~.
\end{equation}

With $\volumeproducts$ and the enthalpy of reaction $\deltaenthaplyreaction$, we compute $\nuprod$, giving the volume change for one joule of reaction as
\begin{equation}
\nuprod = \frac{\volumeproducts}{| \deltaenthaplyreaction |} = 0.00041 \ \left[\frac{cm^3}{J}\right] ~.
\end{equation}

The change in volume of products $\changevolumeproducts$ resulting from a given change in cumulative heat $\deltacumulheatJ$ is then computed as
\begin{equation}
\changevolumeproducts = \nuprod \deltacumulheatJ ~.
\end{equation}

We note that $\deltacumulheatJ$ is expressed in Joules, whereas changes in cumulative heat recorded from isothermal calorimetry $\deltacumulheatHzero$ are expressed in Joules per unit mass of binder. Isothermal calorimetry data thus allow to compute changes in volume of products \emph{per unit mass of binder} with $\changevolumeproducts/\mass_{\mathrm{binder}} = \nuprod (\deltacumulheatHzero)$.

\section{Exponential fits on storage modulus vs. thickness of products data}
\label{sec:p02:appendix_expfits_R2}

\begin{figure}[htbp]
    \centering
    \includegraphics[width=0.95\linewidth]{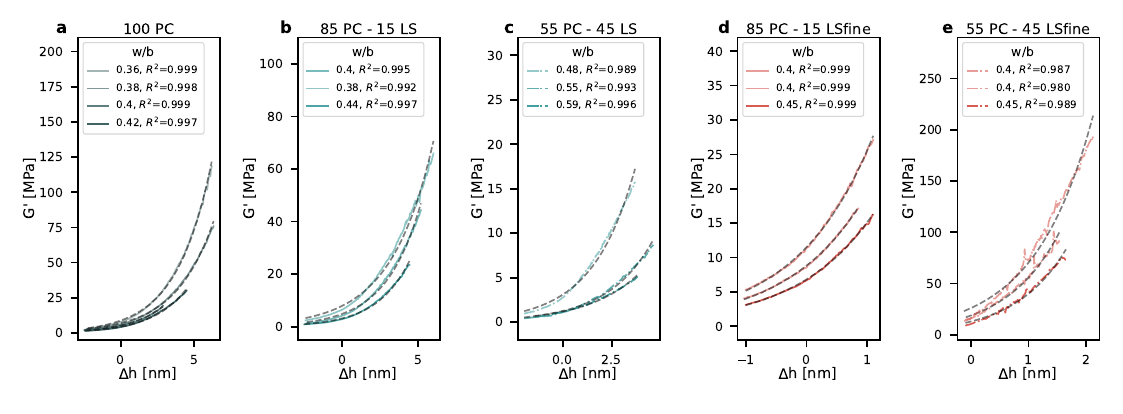}
    \caption{\textit{Exponential fits on $\storagemodulus$ vs. $\deltaproductsthickness$ data. The full lines are the data, the dashed lines are the fits.\textbf{a)} Plain Portland cement paste. \textbf{b)} 85\% Portland cement-15\% coarse Limestone blend. \textbf{c)} 55\% Portland cement-45\% coarse Limestone blend. \textbf{d)} 85\% Portland cement-15\% fine Limestone blend. \textbf{e)} 55\% Portland cement-45\% fine Limestone blend. Compositions given in mass fractions.}}
    \label{fig:p02:expfits}
\end{figure}

We provide exponential fits on $\storagemodulus$ vs. $\deltaproductsthickness$ data for each system investigated, along with $R^2$ values, in Fig.~\ref{fig:p02:expfits}. The $R^2$ values are systematically above 0.98, showing good agreement between the experimental data and the fitted exponential from Eq.~\ref{eq:p02:G_aexp_bdh}.

\section{Dependence of $\storagemodulusref$ on $\solidvolfrac$}
\label{sec:p02:appendix_G0phi}
\begin{figure}[htbp]
    \centering
    \includegraphics[width=0.6\linewidth]{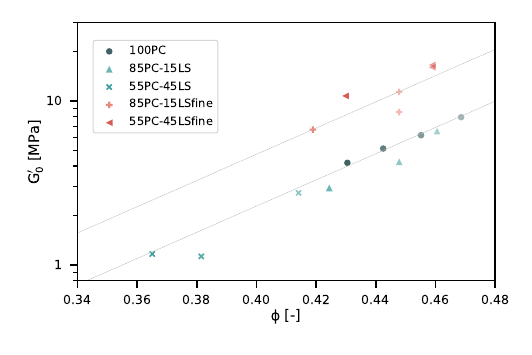}
    \caption{\textit{Dependence of the storage modulus at the onset of the acceleration period $\storagemodulusref$ on the solid volume fraction $\solidvolfrac$. Note the logarithmic scale on the y-axis. The dashed lines serve as guides for the eyes.}}
    \label{fig:p02:G0_phi}
\end{figure}

The storage modulus at the onset of the acceleration period $\storagemodulusref$ increases with solid volume fraction $\solidvolfrac$, as shown in Fig.~\ref{fig:p02:G0_phi}. $\storagemodulusref$ is expected to diverge at the maximum packing fraction. We note that we consider here the initial solid volume fraction, \ie, the volume fraction of unhydrated cement grains. This is because volume fraction changes are either not accessible with the current experimental setup (very early hydration products), or negligible (low amounts $\csh$ and Portlandite forming over the time periods considered here). The small range of $\solidvolfrac$ values makes it difficult to compare different fitting functions. We nevertheless observe that exponential relations do capture the increasing trend, inline with previous studies~\cite{michelStructuralBuildrestInduction2024,mantellatoShiftingFactorNew2020}.

\section{SSA of Portland cement after contact with water}
\label{sec:p02:appendix_wet_ssa}
\begin{figure}[htbp]
    \centering
    \includegraphics[width=0.6\linewidth]{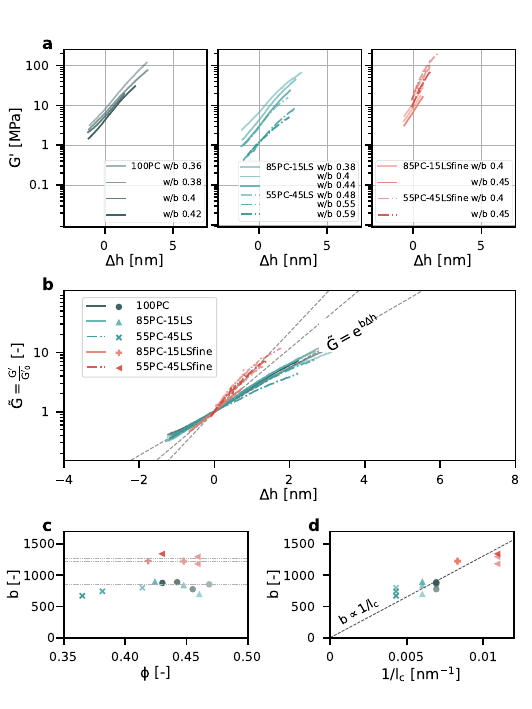}
    \caption{\textit{Considering the $\ssa$ of Portland cement after contact with water (2.2 m$^2$/g over the dry $\ssa$ (1.1 m$^2$/g) \textbf{a)} Storage modulus evolution with changes in thickness of hydration products at different w/b ratios. Left: Portland cement. Middle: blends of Portland cement with coarse limestone. Right: blends of Portland cement with fine limestone. \textbf{b)} Normalized storage modulus vs. changes in thickness of hydration products for blends with and without limestone. \textbf{c)} Exponent of the $\storagemodulusnorm$ vs. $\deltaproductsthickness$ relation as a function of the initial solid volume fraction $\solidvolfrac$ (unhydrated grains). \textbf{d)} The exponent of the $\storagemodulusnorm$ vs. $\deltaproductsthickness$ relation scales linearly with the inverse of the characteristic length, $1/\characteristiclength$.} }
    \label{fig:p02:G_dh_wet_ssa}
\end{figure}

The $\ssa$ of Portland cement increases substantially shortly after contact with water, raising from 1.1 m$^2$/g to 2.2 m$^2$/g after 10 minutes~\cite{mantellatoRelatingEarlyHydration2019}. Considering the $\ssa$ after contact with water over the dry $\ssa$ changes the evolutions of $\storagemodulus$ and $\storagemodulusnorm$ with $\deltaproductsthickness$, see Fig.~\ref{fig:p02:G_dh_wet_ssa}a\&b. For a given system, the exponent of the $\storagemodulusnorm$~vs.~$\deltaproductsthickness$ remains independent of solid volume fraction, see Fig.~\ref{fig:p02:G_dh_wet_ssa}c. Furthermore, when considering the $\ssa$ after contact with water, the exponent $b$ scales linearly with $1/\characteristiclength$, see Fig.~\ref{fig:p02:G_dh_wet_ssa}d.

\newpage %

\end{document}